\documentstyle[aps,prd,epsfig,amstex]{revtex}

\begin{document}

\draft

\wideabs{
\title{Search techniques for gravitational waves from black-hole ringdowns}
\author{Jolien D. E. Creighton}
\address{Theoretical Astrophysics,
         California Institute of Technology,
         Pasadena, California 91125}
\date{28 January 1999}
\maketitle
\begin{abstract}%
Of all the astronomical sources of gravitational radiation, the ringdown
waveform arising from a small perturbation of a spinning black hole is perhaps
the best understood: for the late stages of such a perturbation, the waveform
is simply an exponentially-damped sinusoid.  Searching interferometric
gravitational wave antenna data for these should be relatively easy.  In this
paper, I present the results of a single-filter search for ringdown waveforms
arising from a 50 solar mass black hole with $98\%$ of its maximum spin
angular momentum using data from the Caltech 40-meter prototype
interferometer.  This search illustrates the techniques that may be used in
analyzing data from future kilometer-scale interferometers and describes some
of the difficulties present in the analysis of interferometer data.
Most importantly, it illustrates the use of coincident events in the output
of two independent interferometers (here simulated by 40-meter data at two
different times) to substantially reduce the spurious event rate.
Such coincidences will be essential tools in future gravitational wave
searches in kilometer-scale interferometers.
\end{abstract}
}

\narrowtext

\section{Introduction}
\label{s:intro}

The first kilometer-scale gravitational wave observatories, LIGO~\cite{a:1992}
and VIRGO~\cite{c:1997}, are expected to begin performing searches for
astrophysical sources of gravitational waves in the year 2002.  By this time,
it will be necessary to have data analysis software for on-line searches.
Because the anticipated signals are expected to be weak, near-optimal
strategies for data analysis will be required in order to extract
gravitational wave signals from the interferometer noise.

An optimal technique when the waveform of the signal is known in advance is
the method of matched filtering~\cite{h:1968}.  The statistical properties of
the matched filter may be characterized when the detector noise is stationary
and Gaussian.  However, in order to understand the behavior of the matched
filter for a real detector, it is necessary to apply the method to a
problem of data analysis that resembles, in some way, the expected
non-stationary and non-Gaussian LIGO and VIRGO interferometer data.  In this
paper, I report the results of analyzing data acquired from the Caltech
40-meter prototype interferometer~\cite{40-meter} in November 1994.  The noise
spectrum of the prototype instrument had some of the same characteristics of
the spectra expected in LIGO and VIRGO: at low frequency it was dominated by
seismic ground motion and at high frequency it became photon shot-noise
limited.  However the 40-meter prototype experienced many still
uncharacterized non-Gaussian transient disturbances.  In this respect, the
40-meter prototype is expected to be much worse than the LIGO and VIRGO
interferometers.  Nevertheless, the prototype forms a useful test-bed for
development of robust algorithms that may be used with some degree of
confidence in analyzing the data of future instruments.

The data analysis problem I present is a search for ringdown gravitational
waves emitted from a distorted spinning black hole.  Such systems will form as
a final state in the merger of black-hole binaries.  These ringdown waves are
important candidates for detection by the kilometer-scale interferometers for
sufficiently massive ($50M_\odot\lesssim M\lesssim500M_\odot$) black holes;
for such masses, the ringdown waves lie in the frequency band of greatest LIGO
sensitivity, $\sim30$--$300\,\text{Hz}$.  Flanagan and Hughes~\cite{fh:1998}
have shown that black-hole ringdowns are detectable to distances of
$\sim10\,\text{Mpc}$--$1\,\text{Gpc}$ for the initial LIGO interferometers.
For the Caltech 40-meter prototype, black-hole ringdown waveforms could be
detected in our Galaxy (see the Appendix and Fig.~\ref{f:ring}).  Because the
ringdown waveform is an exponentially-damped sinusoid, the construction of
filters to search for such a signal is simple.  A search for ringdown
waveforms, therefore, provides a useful test of the application of matched
filtering to interferometer data.

Damped sinusoids are also expected to be produced by a variety of spurious
events in the interferometer.  For this reason, one would not conclude that
any particular event recorded by a matched filter necessarily corresponds
to a gravitational wave, no matter how large the event is, unless there is
sufficient corroborating evidence.  Possible types of corroborating evidence
include the presence of a preceding inspiral and merger waveform, which would
be expected for the coalescence of a binary system, and a coincident event
that is detected in another interferometer.  A search for coincident events
in detectors with independent noise is an \emph{essential} technique for
discriminating between instrumental effects (which are unlikely
to produce coincident events) and gravitational wave signals (which are
expected to produce coincident events).  In addition to these methods of
corroboration, a full-scale search for ringdown waveforms would involve
additional vetoes based on environmental monitors and on a catalog of known
instrumental effects.  In this
paper, I will concentrate solely on the use of a coincident strategy to
discriminate between ringdowns arising from spurious instrumental effects and
(simulated) gravitational waves.

The purposes of this paper are (i) to present the data analysis for a search
for black-hole ringdowns in data obtained from the Caltech 40-meter
interferometer, with particular emphasis on the use of coincidences between
two detectors to deal with an excess of false alarms produced by the
non-Gaussian component of the detector noise; and (ii) to present the details
and results of an actual search limited to a single template for a black hole
with a particular mass and spin.  A more general search would use a bank of
filters to cover a range of possible black-hole masses and spins.
Furthermore, a more general ringdown search would incorporate many more
discriminants based on a better characterization of the detector noise.
For this reason, the search presented in this paper should be regarded as
an exploratory search that implements the techniques of optimal filtering
and coincidence detection.

I start with a discussion of the statistical problem of signal extraction for
the anticipated ringdown waveform.  For this statistical analysis, it is
common to make the assumptions that the detector noise is a stationary and
Gaussian random process~\cite{h:1968}; I will make similar assumptions, though
they are not valid for the Caltech 40-meter data.  For this reason, I describe
a technique one can use to deal with non-Gaussian components of the detector
noise, if they are sufficiently infrequent, given two detectors with
independent noise.  I then present the results of an application of the
matched filter data analysis technique to the Caltech 40-meter data and
describe the implications of the results for data analysis strategies.  I
simulated a coincidence search by dividing the 40-meter data into two sets,
each set treated as if it came from a separate instrument with independent
noise.  In the Appendix, I describe some details about the ringdown waveform
and the number of ringdown filters that would be required in order to search
for all possible ringdowns in the bandwidth of the 40-meter prototype.

\section{Detection strategy}
\label{s:detect}

In order to detect the ringdown from a black hole, one must pass the
interferometer output through a receiver that will perform some test of
the hypotheses ``there is a signal present in the data'' ($H_1$) and ``there
is no signal present in the data'' ($H_0$).  In order for the receiver to
conclude that there is a signal present, it constructs some statistic and
compares the statistic to some pre-assigned threshold.  The problem of
reception, then, is two-fold: one must choose some optimal statistic, and one
must select some threshold.  I will explore these problems below.  The
discussion in this section follows the method presented in
Refs.~\cite{h:1968,f:1992}.

In designing the optimal statistic, I will assume, in this section, that the
noise present in the detector is a stationary Gaussian process; this
assumption simplifies the statistical analysis.  However, it is known that the
Caltech 40-meter interferometer data have significant non-stationary and
non-Gaussian noise components, so the optimal statistic will not have the
properties expected of stationary Gaussian noise.  The effect of the
non-stationary and non-Gaussian noise components will be evident in the
observations presented in Sec.~\ref{s:search}.

\subsection{The matched filter}
\label{ss:matched}

Suppose that the detector strain output $h(t)=h_{\text{GW}}(t)+n(t)$ consists
of stationary Gaussian noise $n(t)$ and, under hypothesis $H_1$, a strain
caused by a gravitational wave $h_{\text{GW}}(t)=Aq(t)$ of known form.  The
waveform $q(t)$ is the expected gravitational waveform at some fiducial
distance so that the amplitude $A$ represents the inverse distance of the
source in units of this fiducial distance.  The stationary (colored) Gaussian
noise $n(t)$ can be characterized entirely by its one-sided noise power
spectrum $\langle\tilde{n}(f)\tilde{n}^\ast(f')\rangle
=\frac{1}{2}\,S_h(|f|)\delta(f-f')$ where $\tilde{n}(f)$ is the Fourier
transform of the noise and $\ast$~denotes complex conjugation.  If the noise
is known to be a stationary Gaussian process, then the matched filter
statistic
\begin{equation}
  x = (h\mid q) = \int_{-\infty}^\infty df\,
    \frac{\tilde{h}^\ast(f)\tilde{q}(f)+\tilde{h}(f)\tilde{q}^\ast(f)}
         {S_h(|f|)},
  \label{e:inner_product}
\end{equation}
is the optimal statistic.  That is, the probability of $H_1$ given the
detector output, $P(H_1\mid h)$, depends on the output only through the
matched filter statistic $x$, and the probability is a monotonically
increasing function of $x$.  To decide between the two hypotheses, one selects
some threshold $x_\star$ and accepts hypothesis $H_1$ if $x>x_\star$, otherwise
one accepts hypothesis $H_0$.  For one to choose the threshold in terms of
some desired probability $P(H_1\mid h)$ requires that one know the prior
probability of hypothesis $H_1$.  The prior probability for $H_1$ is usually
subjective; however, one often wants an objective way of choosing the
threshold $x_\star$.  In this case, one usually constructs a threshold based on
the false alarm probability $Q_0$ or the true detection probability $Q_1$.
(The true detection probability is the converse of---i.e., one minus---the
false dismissal probability.)

Two important properties of the matched filter statistic are the false alarm
and true detection probabilities, which are obtained from the probability
distribution for the matched filter under the two hypotheses.  Under
hypothesis $H_0$, the probability distribution for $x$ is a Gaussian
distribution with zero mean and variance $\sigma^2=(q\mid q)$ where the inner
product is obtained from replacing $h$ with $q$ in
Eq.~(\ref{e:inner_product}).  However, when a signal is present with amplitude
$A$, the mean of this distribution is shifted to $A\sigma^2$.  It will be
useful to consider the signal-to-noise ratio $\rho=|x|/\sigma$ (the absolute
value is taken because it is not known whether the signal will have a positive
or negative amplitude).  The false alarm and true detection probabilities are
\begin{equation}
  Q_0 = P(\rho > \rho_\star \mid H_0) = {\rm{erfc}}(\rho_\star/\surd 2)
  \label{e:fa_prob}
\end{equation}
\begin{eqnarray}
  Q_1 &=& P(\rho > \rho_\star \mid H_1) \nonumber\\
      &=& {\textstyle\frac{1}{2}}\,
          {\rm{erfc}}[(\rho_\star-A\sigma)/\surd2]
          + {\textstyle\frac{1}{2}}\,
          {\rm{erfc}}[(\rho_\star+A\sigma)/\surd2]
  \label{e:td_prob}
\end{eqnarray}
where $\rho_\star$ is the signal-to-noise ratio threshold and the complementary
error function is defined by~${\rm{erfc}}(x)=2\pi^{-1/2}\int_x^\infty
e^{-t^2}dt$.  Using these equations, one can choose a threshold for a desired
false alarm or true detection probability.

In the above discussion, I have made the implicit assumption that one knows
the arrival time of the signal.  Since this will not be known in general, it
is necessary to obtain the matched filter statistic $x(t_{\text{arr}})$ as a
function of possible arrival times $t_{\text{arr}}$.  This can be accomplished
by replacing $q(t)$ with $q(t-t_{\text{arr}})$ or, equivalently, by replacing
$\tilde{q}(f)$ with $\tilde{q}(f)e^{2\pi ift_{\text{arr}}}$ in
Eq.~(\ref{e:inner_product}).  Whenever the signal-to-noise ratio exceeds the
threshold, a candidate event has occurred.  Because of correlations in the
time series $x(t_{\text{arr}})$, a single ringdown waveform may cause several
threshold crossings at slightly different (but closely spaced) arrival times;
thus, it is necessary to maximize the signal-to-noise ratio over events with
similar arrival times.

The situation is further complicated if the waveform depends on parameters
other than the time of arrival.  In order to search over these parameters, it
is necessary to construct a bank of filters that span the parameter space.
Thus, if $\{\hat{q}_i(t)\}$ for $i=1,\ldots,N_{\text{filters}}$ is a set of
$N_{\text{filters}}$ filters that cover the parameter space, then one has a
set of matched filter time series $\{x_i(t_{\text{arr}})\}$.

Because there are correlations between the different outputs
$\{x_i(t_{\text{arr}})\}$ at a given arrival time as well as within each time
series itself, it is difficult to estimate the rate of false alarms for the
filter bank.  An overestimate can be obtained by assuming that each filter and
each arrival time is independent.  Then the rate of false alarms is
\begin{equation}
  R_0 = N_{\text{filters}} \Delta^{-1} {\rm{erfc}}(\rho_\star/\surd2),
  \label{e:rate}
\end{equation}
where $\Delta^{-1}$ is the sampling rate of the detector.  A more accurate
estimate could be obtained by a Monte Carlo simulation of the detection
process in the presence of Gaussian noise alone.

The probability of true detection in the presence of a bank of time-series
filters must be estimated by Monte Carlo methods.  However, an underestimate
of this probability can be made by assuming that the detection can only occur
at the correct time and with the correct filter; then the probability given in
Eq.~(\ref{e:td_prob}) can be used.  This estimate is good for signals with
large amplitudes, but it becomes a significant underestimate when the
probability of a false alarm becomes comparable to the probability of true
detection.

A calculation of the number of filters $N_{\text{filters}}$ necessary for a
search for all possible black-hole ringdown waveforms within the bandwidth of
the Caltech 40-meter prototype is given in the Appendix.  For the present
work, however, I will report on the computationally easier problem of a single
filter search of the 40-meter data, which captures most of the issues that
arise due to the non-Gaussian nature of the 40-meter noise.  The results of a
full filter bank search will be reported elsewhere~\cite{c:1998}.

\subsection{Dealing with non-Gaussian noise}
\label{ss:nonGauss}

It is unlikely that all possible sources of non-Gaussian instrumental noise
will be eliminated from the interferometers, and even if most of the remaining
non-Gaussian components can be removed from the data before analysis, there
will likely be some non-Gaussian processes occurring at some (hopefully small)
rate.  It is important to have relatively robust methods for dealing with such
events because they will likely occur more frequently than the desired false
alarm event rate.  I will discuss a very simple method for ``removing'' the
non-Gaussian events based on coincidence between two detectors in a
network.  Such coincidences have long been central to data analysis for
resonant-mass gravitational-wave detectors (see, e.g., Ref.~\cite{k:1977})
and were regarded as essential in the original planning of
LIGO~\cite{v:1989}.  That is why the LIGO network will consist of two
full-sized detectors that will have largely independent noise (since they will
be separated by $3002\,\text{km}$), plus an additional half-sized detector at
one of the two sites.  VIRGO could provide another full-sized detector to the
network, and since it will be located on a different continent, it may have
almost entirely independent noise.

Suppose, for simplicity, that there are only two detectors available, and that
the detectors have the same sensitivity and independent noise.  Also, suppose
that one expects that the arrival times of a signal will agree to within some
time $\pm\Delta\tau$, which will be approximately the light travel time
between the two detectors ($\pm10\,\text{ms}$ between the LIGO sites).  One
can then adopt the following na{\"\i}ve detection strategy: accept an event as
a candidate signal if it occurs with a signal-to-noise ratio greater than
$\rho_\star$ in both detectors, and the times of arrival agree to within
$\pm\Delta\tau$.%
\footnote{Clearly, this is a crude strategy which can be improved by a more
careful likelihood analysis.  Such an analysis has been done by
Finn~\cite{f:1998}, in which the outputs of all detectors are jointly
analyzed.  The na{\"\i}ve strategy might be used as a first stage in a
hierarchical strategy that will combine detector output only after first
identifying candidate signals.}
If the threshold $\rho_\star$ corresponds to a false alarm rate of
$R_0^{\text{single}}$ in each detector, and if the noises in the two detectors
are independent, then the fraction of false alarms that are coincident will be
$R_0^{\text{single}}\times2\Delta\tau$, and the overall rate of coincident
false alarms is
$R_0^{\text{coincident}}=(R_0^{\text{single}})^2\times2\Delta\tau$.  (The
factor of two accounts for the fact that both positive and negative time
delays of up to $\Delta\tau$ are allowed.) In practice, this means that the
threshold should be set at a level such that the false alarm rate in the
individual detectors is
\begin{equation}
  R_0^{\text{single}} = (R_0^{\text{coincident}}/2\Delta\tau)^{1/2}
\end{equation}
for some desired overall coincident false alarm rate.  For example, for the
two full-sized LIGO interferometers (not including the half-sized
interferometer at Hanford) and an overall false alarm rate of one per ten
years, the individual thresholds should be set so that the individual detector
false alarms occur less frequently than one every $1.2$~hours.  Such a false
alarm event rate for the individual detectors may be larger than the rate of
non-Gaussian events; if this is indeed the case, then the
thresholds~$\rho_\star$ can be set to the desired false alarm rates based on
the results of Sec.~\ref{ss:matched}, which assumed Gaussian noise alone.

The overall coincident true detection probability will be the product of the
true detection probabilities from the two interferometers, i.e.,
$Q_1^{\text{coincident}}=(Q_1^{\text{single}})^2$ if the two detectors have
roughly equal true detection probabilities for a given signal.  Thus one
requires a smaller signal-to-noise ratio threshold in each of the two
detectors for a coincident search than for a single detector search in order
to achieve the same true detection probability.  A useful measure of the
effectiveness of a search strategy is the maximum effective observing time
$T_{\text{eff}}=Q_1/R_0$.  This is the amount of time one can observe before
encountering a false alarm, times the efficiency of detecting a signal.  By
multiplying this effective time by the rate of astronomical events, one
obtains the expected number of observed events before a false alarm occurs.
Thus, the rate of astronomical events should be much greater than
$1/T_{\text{eff}}$ in order to cleanly distinguish them from false alarms.
One finds that a coincident search strategy is better---i.e., yields a larger
value of $T_{\text{eff}}$---than a single detector search provided that
$Q_1^{\text{single}}>R_0^{\text{single}}\times2\Delta\tau$.  This inequality
will be satisfied for any realistic search for gravitational waves since it
merely requires that the probability of detecting an event, when present,
exceeds the probability of a false alarm in the short time $2\Delta\tau$.

Unlike the anticipated LIGO and VIRGO detectors, the Caltech 40-meter
prototype is a research and development instrument whose configuration
is frequently changed without taking the time to completely eliminate
or characterize the sources of non-Gaussian noise.
The result is that the Caltech 40-meter data tend to have many non-Gaussian
events and also tend to be non-stationary.  As will be shown below, this means
that the false alarm probability distribution does not agree with the
predictions made in Sec.~\ref{ss:matched}\@.  Nevertheless, if two 40-meter-like
interferometers with independent noise were available, it would still be
possible to achieve a desired overall false alarm rate using the argument
above but where the individual false alarm rate distributions would have to be
measured in order to set the threshold.  Because the required
\emph{individual} false alarm rates $R_0^{\text{single}}$ in LIGO may be on
the order of one per hour (so that the coincident false alarm rate would be on
the order of one per ten years if the two interferometers were separated by
$3002\,{\text{km}}$), it will be easy to determine the necessary threshold
$\rho_\star$ by analyzing several hours of data.  If one was not able
to do a coincident search and if one required a false alarm rate of one per
year, then one might hope to set the correct threshold by analyzing several
years worth of data in order to be sure to set the correct threshold level.
Unfortunately, this could well be insufficient because it is not possible to
screen out gravitational wave signals in order to determine the properties of
the noise alone.  Thus, unless the properties of the detector noise are known
exactly, the rate of false alarms cannot be estimated for a single detector,
but the rate can be estimated for a coincident search using multiple
detectors.

\section{Single filter search}
\label{s:search}

Having reviewed a possible detection strategy and stated the expected
distribution of the detection statistic in the presence of stationary Gaussian
noise, I now examine the results of applying the detection strategy to real
interferometer data.  In November 1994, the Caltech 40-meter prototype
interferometer was used to collect approximately 46 hours of data.  In my
analysis of these data, I implemented the detection strategy discussed in the
previous section (for a single filter only) using code that is provided in the
{\sc{grasp}} data analysis software package~\cite{a:1997}.  I also made
extensive use of the Numerical Recipes library~\cite{ptvf:1992} and the
{\sc{fftw}} Fourier transform routine~\cite{fj:1997}.

I analyzed the data by correlating them with a single filter function
corresponding to the fundamental quadrupole quasinormal mode of a Kerr black
hole with a mass~$M=50M_\odot$ and a spin $\hat{a}=98\%$ of the maximum spin.
This mode is a damped sinusoid with a central frequency of
$f\simeq520\,{\text{Hz}}$ and a quality factor of~$Q\simeq12$ (see the
Appendix).  The central frequency of the ringdown is within the frequency band
of the instrument: between approximately $100$ and $5000\,\text{Hz}$ (see
Fig.~\ref{f:noise}).  The filter was constructed in the frequency domain as
$\tilde{q}(f)/S_h(|f|)$ [the ``over-whitened'' ringdown waveform required in
Eq.~(\ref{e:inner_product})]; the filter was then truncated in the time domain
so that it had a total duration of $\tau_{\text{filter}}=1.66\,{\text{s}}$.
The truncation is necessary because the power spectrum $S_h(|f|)$ possesses
narrow line features that would normally cause the filter to have a very long
impulse response; by truncating the filter, these line spikes are not so well
resolved, but the impulse response becomes much shorter.%
\footnote{The additional filter duration could be reduced significantly if
the line features were removed in some pre-processing phase.  It is likely
that this will be done in LIGO prior to the filtering of the data, so the
ringdown filters may not need to be so long for a search in LIGO data.}
It is found that the output from this filter is in good agreement with the
output of a filter with a much longer $\tau_{\text{filter}}$; the loss in
signal-to-noise ratio caused by the somewhat sub-optimal filter is less
than~$10\%$.

The data were filtered in the frequency domain in segments of
$6.64\,{\text{s}}$, or $2^{16}$~samples at the instrumental sampling rate of
$\Delta^{-1}\simeq9.868\,{\text{kHz}}$.  To remove the effect of the
correlation wrap-around, the last $\tau_{\text{filter}}=1.66\,{\text{s}}$
(corresponding to the impulse response of the filter) of the correlation output
was dropped for each segment; the data segments were overlapped by the same
amount to compensate for the dropped output.  Only the data from the times in
which the instrument was in lock were used, and data segments in which there
were outliers (defined as any segment in which there was an interferometer
sample that was more than five times the sample standard deviation for the
segment, or too many samples more than three times the sample standard
deviation for the segment) were discarded.%
\footnote{These outliers are certainly due to noise transients and cannot be
regarded as black hole ringdowns: for a black hole ringdown to have produced
such a large amplitude spike, it would have to have been in the Solar
neighborhood.  In a full-scale search, events such as these would not be
dismissed outright; rather, they would be collected and carefully examined to
establish that they are instrumental artifacts.  However, for this analysis it
is easy to see from the time series that these outliers are not ringdown
waveforms so I have excluded them from this ringdown search.}
The power spectrum of the instrument that was used to construct the filter was
obtained by averaging the spectra of 8 segments of data before and after the
start of each segment.

As a stepping-stone towards examining a coincidence-style search for
gravitational waves, I describe the results of a single-detector search
below. These results show the necessity of devising methods for reducing
the false alarm rate caused by non-Gaussian nature of the noise.
A coincident event in an independent detector is the most immediate
corroboration available (though further corroboration would also be
required before one concluded that an event was a gravitational wave).
A description of a simulated coincidence search with 40-meter data
follows the single detector results.

\subsection{Single detector results}

\begin{figure}[p]
\epsfig{file=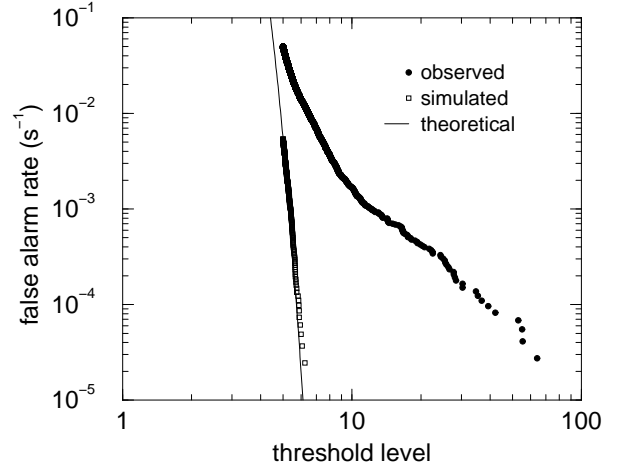,width=0.9\linewidth}
\caption{The false alarm rate for a \emph{single} detector as a function of
threshold level $\rho_\star$.  The filled circles represent the observed false
alarm event rate from filtering the Caltech 40-meter prototype interferometer
data while the open squares represent the false alarm event rate obtained by
filtering simulated stationary Gaussian noise.  The solid line is the
theoretical rate for stationary Gaussian noise given by Eq.~(\ref{e:rate})
with $N_{\text{filters}}=1$ filters.}
\label{f:fa_rate}
\end{figure}

After the filtering process, I was left with 20~hours of filter output.  From
this output, I produced a list of ``events'' for which the filter output
exceeded a signal-to-noise ratio threshold of~5.  I next produced a ``distinct
event'' list by maximizing the signal-to-noise ratios over time delays of
$\pm\tau_{\text{filter}}=\pm1.66\,{\text{s}}$ corresponding to the impulse
response time of the filter.  A single signal (or noise transient) may cause
the filter output to exceed the threshold many times.  This maximization
accounts for the time-correlations in the filter output caused by the finite
impulse response of the filter; thus, any ringdown waveform will cause only
one distinct event to be recorded.  I then estimated an event rate as a
function of signal-to-noise ratio greater than~5.  Reasonable estimates of
black-hole ringdown event rates in our our Galaxy and its environs indicates
that it is almost certain that there were no gravitational wave signals of
this brightness in the $46$~hours of data; thus this event rate corresponds to
the measured false alarm rate.  I plot this false alarm rate as a function of
signal-to-noise ratio threshold in Fig.~\ref{f:fa_rate}.  In addition to the
false alarm rate observed from the interferometer data, I plot the false alarm
rate expected from the same filtering process for simulated stationary
Gaussian noise.  Figure~\ref{f:fa_rate} shows that, for a desired false alarm
rate, one would have to set a signal-to-noise ratio threshold greater than the
threshold that would be anticipated from an analysis of stationary Gaussian
noise.  The departure of the false alarm curve from the expected curve
assuming stationary Gaussian noise is often seen in real detector data for
sufficiently low false alarm rates; however, here it seems that the stationary
and Gaussian assumptions are poor even for relatively large false alarm rates.

It should be emphasized that a large number of false alarms is not unexpected
when using a ringdown filter because many kinds of non-Gaussian events could
produce detector output that resembles an exponentially-damped sinusoid.  For
example, the ringdown filter that I used, with a central frequency of
$520\,\text{Hz}$, could be excited by beating between two lines such as the
$500\,\text{Hz}$ and the $540\,\text{Hz}$ lines seen in Fig.~\ref{f:noise};
this effect could become important if a large amount of power were present in
the lines.  A possible remedy would be to remove these lines from the data
before filtering it, but it may also be possible to design a discriminator
that would reject these events.

\begin{figure}[p]
\epsfig{file=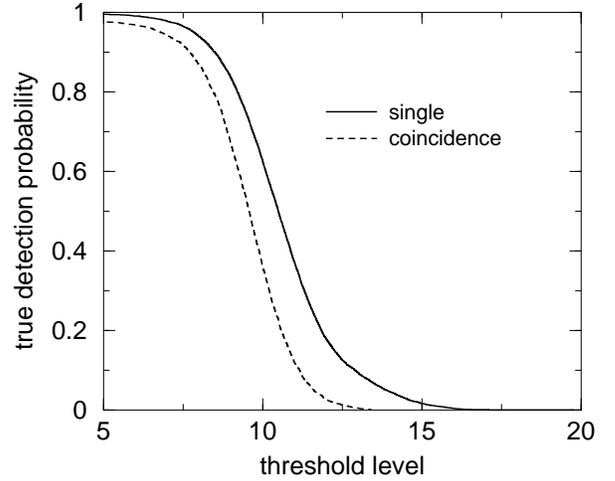,width=0.9\linewidth}
\caption{The true detection probabilities as a function of threshold level
$\rho_\star$ for a simulated black-hole ringdown with mass $50\,M_\odot$ and
$98\%$ of maximum spin at a distance of $50\,{\text{kpc}}$ and a gravitational
wave luminosity equal to $1\%$ of the black hole mass.  The solid line
was measured for injected signals in a single detector search, while the
dotted line was measured for simultaneously injected signals in a simulated
two-detector coincidence search.}
\label{f:td_prob_LMC}
\end{figure}

The true detection probability of the filtering process can also be estimated,
but one requires some assumption of a potential source.  For the present
purposes, suppose one is interested in detecting a typical black hole ringdown
occurring at a distance of the large Magellanic cloud (LMC)
($50\,{\text{kpc}}$ away) that emits $1\%$ of the black hole mass in ringdown
gravitational waves.  Suppose also that one is using the correct filter, i.e.,
the black hole has the same mass and spin that I assumed when constructing the
filter above.  Then the true detection probability is the fraction of times
that an injected signal will be detected by the filtering process (within
$\pm1\,{\text{ms}}$) at a given signal-to-noise ratio threshold.  I injected
signals into the data with random Poisson-distributed times at a mean rate of
1 per 10 seconds; only those injections in data that was actually filtered
were counted.  I plot this true detection probability as a function of
signal-to-noise ratio in Fig.~\ref{f:td_prob_LMC}.

For a given signal, a specification of the desired true detection probability
yields a threshold level which can then be used to compute the false alarm
rate.  For example, if one wanted to detect the above signal with a false
dismissal probability of less than $50\%$, then, from
Fig.~\ref{f:td_prob_LMC}, one would be able to set a threshold signal-to-noise
ratio as high as $\rho_\star\simeq11$.  From Fig.~\ref{f:fa_rate}, one finds
that the false alarm rate will be approximately 4 per hour for this threshold.
To reduce this high rate of false alarms, or to improve the true detection
probability for a given false alarm rate, it is necessary to develop methods
to discriminate between real astronomical events and spurious noise events.
The simplest such method is a coincidence experiment.  I now describe an
application of the coincident detection method of Sec.~\ref{ss:nonGauss} to
the 40-meter data from two separate runs where I pretend that the runs were
simultaneous output of two interferometers.

\subsection{Simulated coincident detection}

To simulate the effect of a coincidence strategy of detection, I used the
following procedure:  I divided the interferometer data into two streams of
(roughly) equal length, and I filtered each of these two streams to obtain two
lists of events where the ``times'' of these events were computed as if the
two streams began at the same instant.  The amount of filtered output was
about 9~hours for each stream.  Before maximizing these two streams over the
impulse response time of the filter
($\pm\tau_{\text{filter}}=1.66\,\text{s}$), I constructed a list of coincident
events, i.e., events for which the arrival times agree to within
$\pm\Delta\tau=10\,{\text{ms}}$---thus, the two streams of data are treated as
if they were produced by two independent 40-meter interferometers running
simultaneously and separated by $3000\,{\text{km}}$.  The signal-to-noise
ratio of each coincident event is taken to be $\rho=\min(\rho_1,\rho_2)$ where
$\rho_1$ was the signal-to-noise ratio recorded in the first interferometer
and $\rho_2$ was the signal-to-noise ratio recorded in the second
interferometer.  The resulting list of events was then maximized over the
minimum signal-to-noise ratio $\rho$ for delays up to the impulse response of
the filter ($\pm\tau_{\text{filter}}=1.66\,\text{s}$) to produce a list of
distinct, coincident events.

The number of coincident false alarm events expected is given by
$N_0^{\text{coincident}}=(R_0^{\text{single}})^2\times2\Delta\tau\times
T_{\text{obs}}$, where $T_{\text{obs}}=9\,{\text{h}}$ is the observation time
in each detector and $R_0^{\text{single}}$ is the rate of false alarms
expected in each detector for a desired signal-to-noise ratio threshold.  For
a threshold signal-to-noise ratio of~5, one expects
$N_0^{\text{coincident}}(\rho_\star=5)\simeq2$ events based on an estimate of
$R_0^{\text{single}}(\rho_\star=5)\simeq5\times10^{-2}\,{\text{s}}^{-1}$
obtained from Fig.~\ref{f:fa_rate}.  However, I obtained $18$~coincident
events for a threshold $\rho_\star=5$.  The origin of this problem is the
following: Because the impulse response time of the filter is
$\tau_{\text{filter}}=1.66\,\text{s}$, a loud non-Gaussian noise burst can
potentially corrupt up to $1.66\,\text{s}$ of filter output even if the noise
burst duration is far shorter.  Thus, although I demand that a coincidence of
two events requires that the two events occur within a time
$\pm\Delta\tau=10\,\text{ms}$, it is possible that two short noise bursts in
different detectors and separated in time by up to $2\tau_{\text{filter}}$
could produce a coincident event.%
\footnote{Such occurrences would be easily vetoed by a simple examination of
the filter output time series \cite{b:prep}.  The burst responses will be
much longer than would be expected for a ringdown since a ringdown response
will be very short (a few ms).}
This thesis is supported by the following observation: If the two
filter outputs are maximized over the filter duration
$\pm\tau_{\text{filter}}$ \emph{before} searching for coincidences, then I
observe only $3$~coincidences, which is in good agreement with the expected
number ($2$~events).  (However, it is preferable to apply discriminators, such
as a coincidence requirement, prior to maximization.  This is because it is
possible that a filter output will contain both a signal and a louder noise
event; a discriminator applied after maximization will reject the entire
filter output based on the loud noise event, and will miss the signal, while a
discriminator applied before maximization will remove the noise event but
preserve the signal.)

Because of the finite impulse response time of the filter, a coincidence
strategy may not fully achieve the expected reduction in false alarm rate for
coincident times much less than the impulse response time unless additional
vetoes are used.  Nevertheless, the coincident detection strategy already
provides a tremendous reduction in the false alarm event rate.  At a
signal-to-noise ratio threshold of 5, the event rate is about 2 per hour; the
true detection probability for my model ringdown would be almost unity at this
threshold.  Furthermore, for a signal-to-noise ratio threshold of 7, no
coincident events were observed and thus the false alarm rate is likely less
than about 1 per 4 hours.  By injecting signals with the same arrival time
into the two data streams, one can compute the true detection probability
for a coincidence search (see Fig.~\ref{f:td_prob_LMC}).  For a threshold
of 7, the observed true detection probability is
$Q_1^{\text{coincidence}}=94\%$.  [One would estimate that this probability
would be $Q_1^{\text{coincidence}}=(Q_1^{\text{single}})^2=96\%$ based on
the observed value of $Q_1^{\text{single}}=98\%$ in Fig.~\ref{f:td_prob_LMC}
for a threshold of 7.]  Thus, for a given false alarm rate, a much lower
threshold can be used for the coincident threshold than for the single
interferometer, and the false dismissal probability is greatly reduced.

\section{Conclusions}
\label{s:conclusions}

Black-hole ringdowns are a promising source of gravitational radiation for
detection in the kilometer-scale interferometric gravitational wave detectors.
The method of matched filtering is a well-known technique for data analysis
and can easily be applied for searches for black-hole ringdown; the
computational burden will not be great.  Our understanding of the performance
of the matched filter is largely based on statistical arguments that assume
that the detector noise is stationary and Gaussian.  For real detectors, these
assumptions will not be true at some level; for the data collected by the
Caltech 40-meter prototype interferometer in November 1994, they are
relatively poor assumptions.

I have analyzed 20~hours of data obtained from the Caltech 40-meter prototype
interferometer in November 1994 using a matched filter for a black-hole
ringdown.  The sensitivity of the instrument is such that a typical black hole
of mass $\sim\!50M_\odot$ that emits $\sim\!1\%$ of its energy as
gravitational waves should be detectable as far out as the LMC\@.  However,
the presence of non-stationary and non-Gaussian noise components creates a
high rate of false alarms in the single detector.  These effects can be dealt
with, if one has two separated interferometers with independent noise, by
considering a coincidence detection strategy.  To simulate a coincidence
study, I used the first half of the 40-meter data as if it came from one
detector, and the second half as if it came from a second, independent
detector.  The false alarm rate for a given threshold is significantly reduced
(though still not to the level expected from Gaussian noise), while the false
dismissal probability at a desired false alarm rate is greatly reduced for a
simulated source in the LMC.

The coincidence detection is a simple example of a veto that can be used, if
two detectors are available, to discriminate between events that are caused by
detector noise and events that are caused by gravitational waves.  Because
it is possible that two interferometers, though widely separated, may still have
correlated noise, it is necessary to supplement a coincidence veto with other
vetoes based on environmental monitors at the two sites.  A catalog of
peculiar ``waveforms'' produced by vagueries of the interferometers must be
generated and events resembling those ``waveforms'' may need to be vetoed.
Other more elaborate vetoes can also be considered: An example would be a
parameter-space veto that would make sure that the candidate produces the
expected reduction in signal-to-noise for filters with slightly mis-matched
central frequencies and quality factor.  Such a veto would discriminate
between a high amplitude noise spike with a very large quality factor that
would trigger all the filters in a filter bank, and a gravitational wave from
a ringing black hole with a low quality factor that would only trigger filters
with roughly the same quality factor.  Because I performed a single filter
search, I did not assess the ability of such a veto to ``remove'' non-Gaussian
events.  A multi-filter analysis of the 40-meter data is presently being
conducted~\cite{c:1998}.

One needs to be somewhat cautious about applying detection vetoes constructed
for simulated ringdown signals.  Although the signal from the ringdown of a
black hole is known exactly for late times, at an earlier time in the
ringdown, when the amplitudes are larger and linear perturbation theory is
less reliable, the signal is not so well known.  This fact does not prevent a
matched filter from detecting the late time ringdown waveform, but
too-sensitive a veto may reject a real event if the early time waveform is not
accurately modeled.  Furthermore, exponentially decaying (or growing)
sinusoids may model many other potential sources of gravitational radiation;
one would prefer not to dismiss these sources by focusing too closely on the
black-hole ringdown problem.  In this sense, the coincidence veto seems
promising for a first pass through the data because it only assumes that the
source of the signal is external to the instrument, and that the two
instruments have independent internal noise.

\acknowledgments
I would like to thank the LIGO project for kindly making the Caltech 40-meter
prototype data available for this analysis.  I would also like to thank
B.\@ Allen, B.\@ Barish, J.\@ Blackburn, P.\@ Brady, A.\@ Lazzarini,
F.\@ Raab, K.\@ Thorne, R.\@ Weiss, and S.\@ Whitcomb for their input.
This work was supported by an NSERC fellowship and by NSF grant PHY-9424337
and NASA grant NAG5-6840.

\appendix

\section*{Black-hole ringdown}
\label{s:ringdown}

\subsection{Ringdown waveform}

The ringdown waveform of a distorted spinning black hole can be obtained from
the Teukolsky equation~\cite{t:1973} for small distortions.  The radiative
part of the perturbation produces a metric perturbation at large distances
that can be expressed as an exponentially-damped sinusoid~\cite{t:1974} for
the fundamental quadrupole mode (which will dominate the radiation at late
times).  The central frequency $f$ and quality factor $Q$ depend on the mass
$M$ and the spin $S=\hat{a}GM^2/c$ (where $G$ is Newton's constant and $c$ is
the speed of light) of the perturbed Kerr black hole; they are well
approximated by the analytic fit found by Echeverria~\cite{e:1989} to
Teukolsky-formalism calculations by Leaver~\cite{l:1985}:
\begin{equation}\label{e:echeverria_fit_a}
  f \simeq 32\,\text{kHz}\times[1 - 0.63(1 - \hat{a})^{3/10}]
    \biggl(\frac{M_\odot}{M}\biggr)
\end{equation}
\begin{equation}
  Q \simeq 2(1 - \hat{a})^{-9/20}.
\end{equation}
The dimensionless spin parameter $\hat{a}$ lies between zero (Schwarzschild
limit) and unity (extreme Kerr limit), so the quality factor is greater than
2.

The metric perturbation caused by the quasinormal ringing of a Kerr black hole
will cause a relative strain $h_{\text{GW}}(t)$ in the arms of an
interferometric gravitational wave detector; the response of the detector to a
gravitational wave is given in Ref.~\cite{t:1987}.  The response depends on
the sky position of the black hole and on the relative orientation of the spin
axis of the black hole to the zenith of the detector.  The average strain
produced by a ringing black hole at some fixed distance can be obtained by rms
averaging over the various angles; the result is
\begin{equation}
  h_{\text{GW (angle averaged)}}(t) = A q(t),
\end{equation}
where $q(t)$ is an exponentially-damped sinusoid%
\footnote{It is found that the initial phase of the sinusoid, at $t=0$, has
little effect on the results of the data analysis reported here.  Ideally, the
issue of the initial phase would be removed by matching the ringdown waveform
to the proceeding merger waveform.  The factor of $(2\pi)^{1/2}$ is introduced
for convenience.}
\begin{equation}
  q(t) = \left\{
  \begin{array}{ll}
    (2\pi)^{1/2}e^{-\pi ft/Q} \cos(2\pi ft) & \text{for $t\ge0$} \\
    0 & \text{for $t<0$.}
  \end{array} \right.
\end{equation}
The amplitude $A$ depends on the strength of the perturbation and on the
distance to the black hole.  For a perturbation of a black hole at distance
$r$ that radiates a fraction $\epsilon$ of the total mass-energy of the hole,
the amplitude is
\begin{eqnarray}
  A &\simeq& 2.415\times10^{-21} Q^{-1/2}[1 - 0.63(1 - \hat{a})^{3/10}]^{-1/2}
      \nonumber\\
    &&\quad \times \biggl(\frac{\text{Mpc}}{r}\biggr)
      \biggl(\frac{M}{M_\odot}\biggr)
      \biggl(\frac{\epsilon}{0.01}\biggr)^{1/2}.
\end{eqnarray}
The quality factor measures (roughly) the number of cycles in the waveform, so
the amplitude decreases with increasing quality factor for a fixed energy
emitted.

The signal-to-noise ratio of a ringdown waveform is $A\sigma$ where $\sigma$
is the signal-to-noise ratio that would be observed if the ringdown strain
$q(t)$ were produced.  This is
\begin{equation}
  \sigma^2 \approx \frac{2}{S_h(f)}\int_0^\infty q^2(t) dt
           \approx \frac{Q}{f S_h(f)}\quad (\mbox{large $Q$})
\end{equation}
where $S_h(f)$ is the one-sided noise power spectrum of the detector at the
central frequency $f$ [cf. Eq.~(\ref{e:inner_product}) with $h(t)$ replaced by
$q(t)$, which is nearly monochromatic for large~$Q$].  Thus, $A\sigma\approx
AQ^{1/2}/h_{\text{rms}}(f)$ where $h_{\text{rms}}(f)=[f S_h(f)]^{1/2}$ is the
rms noise strain in the detector.

For example, a black hole of mass $50\,M_\odot$ and~$\hat{a}=0.98$ produces a
ringdown with a central frequency~$f\simeq520\,{\text{Hz}}$ and quality
factor~$Q\simeq12$.  At a distance of $10\,{\text{kpc}}$, the typical ringdown
strain on an interferometer would be $4\times10^{-18}$ if $1\%$ of the mass of
the black hole were radiated in the ringdown waves.  For the 40-meter
prototype interferometer, which has a typical rms strain noise level of
$3\times10^{-19}$ at $520\,{\text{Hz}}$, the expected signal-to-noise ratio
would be about~$50$.  At a distance of the LMC ($50\,{\text{kpc}}$), the same
ringdown would produce a signal-to-noise ratio of about~$10$.  For the
detectability of ringdown waveforms in the LIGO interferometer, see
Ref.~\cite{fh:1998}.

\subsection{Number of filters required}

Since the different possible waveforms depend on the mass and spin of the
black hole (or equivalently on the central frequency and the quality factor of
the damped sinusoid), and since these parameters are continuous, it is
necessary to choose a discrete set or bank of waveforms to form a ``mesh''
that covers the parameter space sufficiently finely.  By ``sufficiently
finely,'' I mean that the degradation in the signal-to-noise ratio due to
having a filter with slightly incorrect parameters should be small.  In
addition to these parameters, the ringdown is also described in terms of its
start time and initial phase.  These parameters are closely related: a
ringdown will correlate strongly with another ringdown with a different phase
and a start time shifted by up to one quarter of a cycle.  Thus, by using a
ringdown template with an incorrect initial phase, one will make an error in
arrival time of less than one quarter of a period and will reduce the observed
signal-to-noise ratio by less than the amount that would be accumulated in
this quarter cycle.  For a high quality-factor ringdown, the signal-to-noise
ratio loss will be small.  Since one will maximize over arrival times, only
the central frequency and the quality factor need be considered when
constructing a mesh of ringdown templates.

\begin{figure}[p]
\begin{center}
\epsfig{file=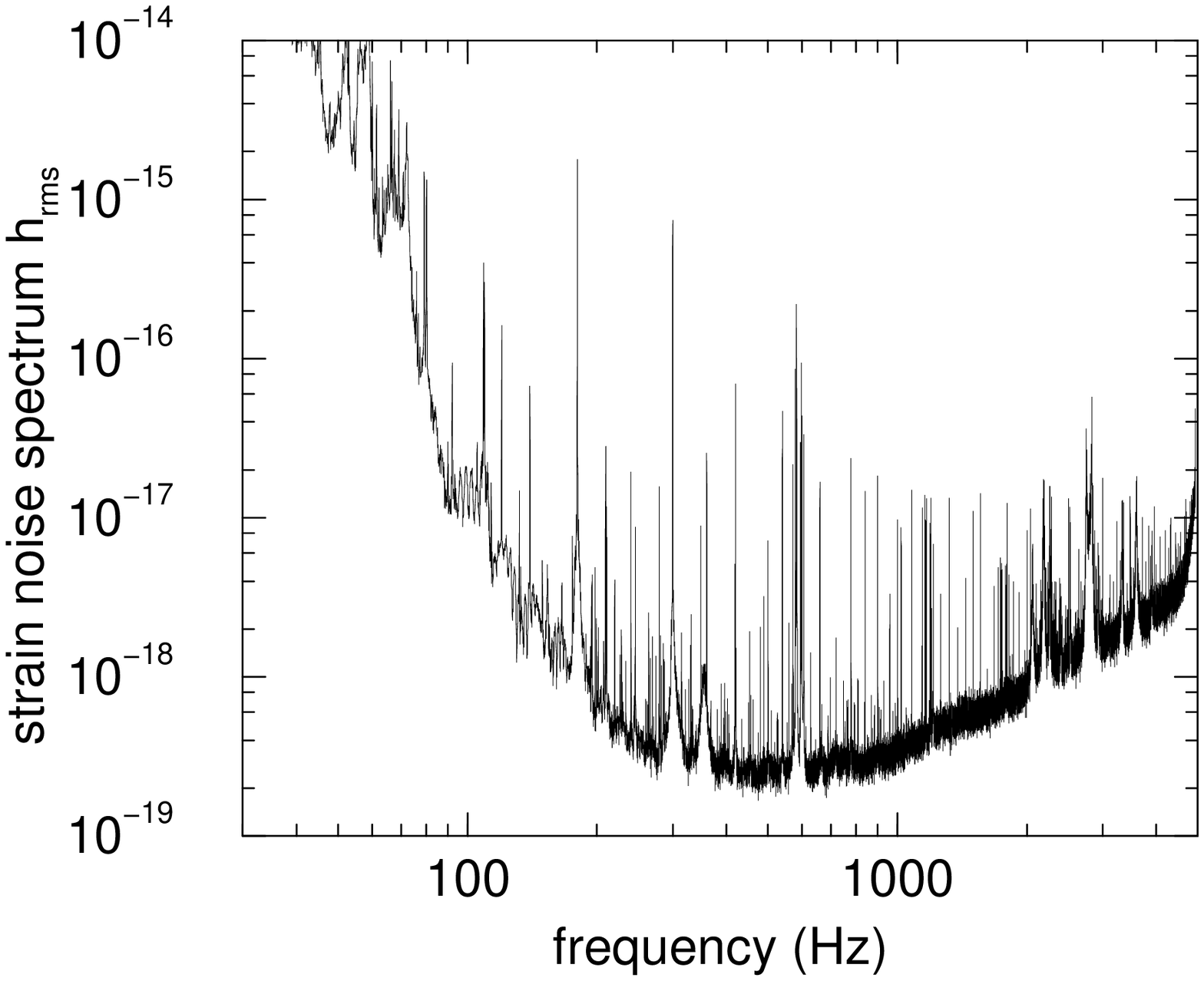,width=0.9\linewidth}\\*
\epsfig{file=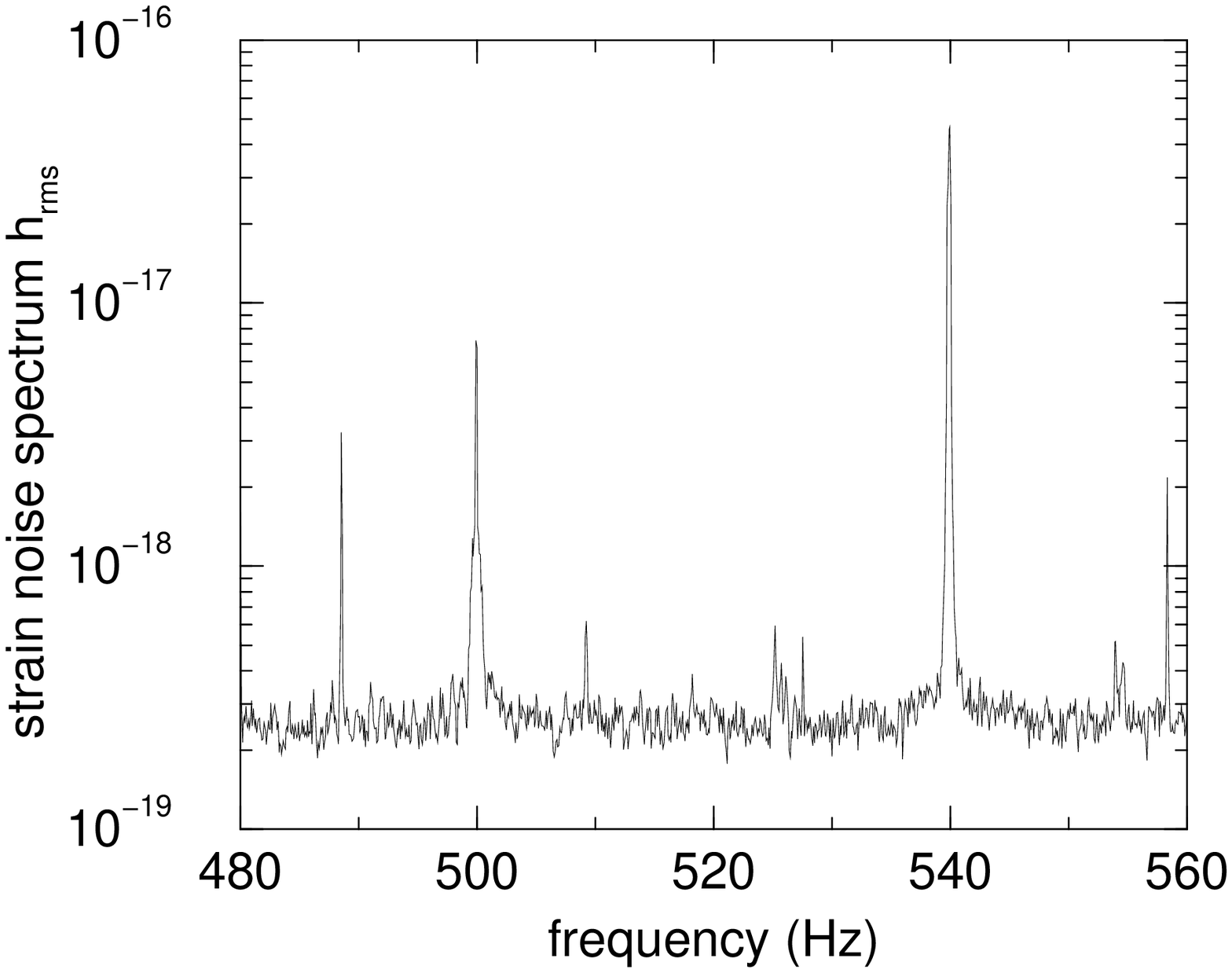,width=0.9\linewidth}
\end{center}
\caption{The rms strain sensitivity $h_{\text{rms}}(f)=[f S_h(f)]^{1/2}$ of
the Caltech 40-meter prototype interferometer in November 1994.}
\label{f:noise}
\end{figure}

In order to estimate how close the templates must be, I follow the procedure
of Owen~\cite{o:1996} in defining a distance function
$ds^2_{ij}=1-(\hat{q}_i\mid\hat{q}_j)$ corresponding to the mismatch between
the two filters $\hat{q}_i$ and~$\hat{q}_j$ that are normalized so that
$(\hat{q}_i\mid\hat{q}_i)=1$.  In the continuum limit, this interval can be
written in terms of a metric as~$ds^2=g_{\alpha\beta}dx^\alpha dx^\beta$ where
$x^\alpha=(f,Q)^\alpha$ is a coordinate on the two dimensional parameter
space.  The metric can be evaluated, in the continuum limit, by
$g_{\alpha\beta}=-\frac{1}{2}(\hat{q}\mid\partial_\alpha\partial_\beta\hat{q})$
where $\partial_\alpha$ is a partial derivative with respect to~$x^\alpha$.
One can show that the mismatch between a filter with frequency and quality
factor $(f,Q)$ and a filter with frequency and quality factor $(f+df,Q+dQ)$ is
\begin{equation}
  ds^2 \approx \frac{1}{8} \frac{dQ^2}{Q^2}
    - \frac{1}{4} \frac{d Q}{Q} \frac{d f}{f}
    + Q^2\,\frac{d f^2}{f^2}.
  \label{e:metric}
\end{equation}
In deriving this equation, I have assumed that the noise power spectrum is
approximately constant over the frequency band of the two filters combined,
and I have kept only the dominant term in the metric coefficients in the limit
of high quality factor.  The minimum number of filters, $N_{\text{filters}}$,
required to cover the parameter space such that there is no point that is a
distance larger than $ds^2_{\text{max}}=(1-\text{``minimal
match''})$~\cite{o:1996} from the nearest template can be found by integrating
the volume element $\surd\det g_{\alpha\beta}$ over the parameter space.  For
the metric of Eq.~(\ref{e:metric}) and the parameter ranges $0\le Q\le
Q_{\text{max}}$ and $f_{\text{min}}\le f\le f_{\text{max}}$, one finds
\begin{eqnarray}
  N_{\text{filters}} &\approx&
    \frac{1}{4\surd2}
    (d s^2_{\text{max}})^{-1}
    Q_{\text{max}}
    \ln(f_{\text{max}}/f_{\text{min}}) \nonumber\\
    &\simeq& 460
    \biggl(\frac{d s^2_{\text{max}}}{0.03}\biggr)^{-1}
    \biggl(\frac{Q_{\text{max}}}{20}\biggr) \nonumber\\
    && \times \biggl\{ 1 + \frac{1}{\log50} \biggl[
    \log\Bigl(\frac{f_{\text{max}}}{5\,{\text{kHz}}}\Bigr)
    - \log\Bigl(\frac{f_{\text{min}}}{100\,{\text{Hz}}}\Bigr)
    \biggr] \biggr\}.
\end{eqnarray}
A value of $ds^2_{\text{max}}=3\%$ is needed in order to ensure that there is
a loss of no more than $10\%$ of the expected event rate due to a mismatched
template~\cite{o:1996}.  The frequency ranges given here are suitable ranges
for the Caltech 40-meter prototype interferometer (see Fig.~\ref{f:noise});
for LIGO and VIRGO, one would choose a different range.

\clearpage

\begin{figure}[p]
\epsfig{file=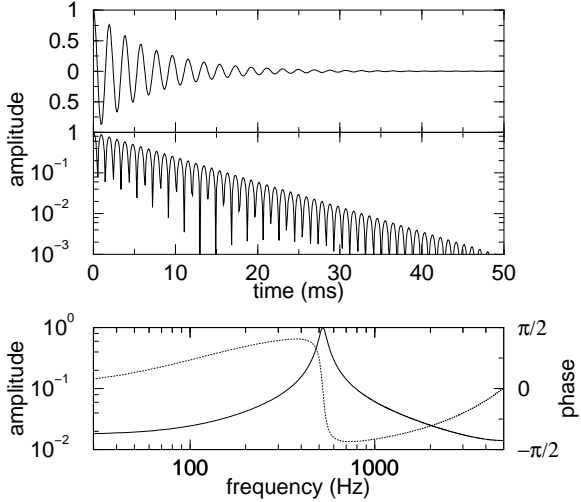,width=0.9\linewidth}
\caption{Time and frequency representations of the ringdown waveform used in
this analysis.  This ringdown has a central frequency of $520\,\text{Hz}$ and
a quality factor of $12$, which corresponds to the fundamental quadrupole
quasinormal mode of a black hole with mass $50\,M_\odot$ and $98\%$ of the
maximum spin.}
\label{f:ring}
\end{figure}

\end{document}